\documentclass[12pt,prd,aps,amssymb,amsmath,tightenlines,a4paper,nofootinbib,superscriptaddress]{revtex4}
\def\bea{\begin{eqnarray}}
\def\eea{\end{eqnarray}}
\def\beq{\begin{equation}}
\def\eeq{\end{equation}}
\def\half{\frac{1}{2}}

\def\vf{\varphi}
\def\ve{\varepsilon}
\def\w{\omega}
\def\p{\partial}
\usepackage[usenames]{color}
\usepackage{graphicx}
\begin{document}
\title{Analytic Solution for $PT$-Symmetric Volume Gratings}

\author{Mykola Kulishov\footnote{email: mykolak@htaphotomask.com}}
\affiliation{HTA Photomask, 1605 Remuda Lane, San Jose, CA 95112, USA}
\author{H.~F.~Jones}
\affiliation{Physics Department, Imperial College, London, SW7 2BZ, UK}
\author{Bernard Kress}
\affiliation{Google, 1600 Amphitheatre Parkway, Mountain View, CA 94043, USA}

\begin{abstract}
We study the diffraction produced by a $PT$-symmetric volume Bragg grating that combines
modulation of refractive index and gain/loss of the same periodicity with a quarter-period shift between them.
Such a complex grating has a directional coupling between the different diffraction orders, which allows us to
find an analytic solution for the first three orders of the full Maxwell equations without resorting to the paraxial
approximation. This is important, because only with the full equations can the boundary conditions, allowing for reflections,
be properly implemented. Using our solution we analyze the properties of such a grating in a wide variety of configurations.
\end{abstract}

\maketitle
\section{Introduction}
Relatively recently it has been discovered that light propagation in an artificial meta-material can be strongly modified, to the extent that this material can become one-way invisible by controlling the Parity-Time ($PT$)-symmetry. Such unidirectional invisibility has been predicted \cite{Pal} for diffraction on a complex refractive index perturbation profile: $\Delta\tilde{n}=\Delta n_0 \exp(2\pi j z/\Lambda)$, which can be realized in practice as the combination of an index grating (real grating) and a balanced gain/loss grating (imaginary grating) using the Euler relation  $\exp(2\pi j z/\Lambda)=\cos(2\pi z/\Lambda)+j\sin(2\pi z/\Lambda)$. It has been shown in the case of a one-dimensional $PT$ symmetric grating that when a beam of light is incident on one side of such a meta-material it is transmitted without any reflection, absorption or phase modulation, which amounts to invisibility of the medium \cite{Pal, MK}.	

$PT$-symmetric gratings have been extensively studied in one-dimensional structures like waveguides \cite{Pal}-\cite{HFJ}, whereas only a few papers \cite{MVB, MKBK, Feng} have addressed diffraction on $PT$-symmetric gratings in free-space configuration or two-dimensional
geometries, as in the case of computer-generated holograms. In these publications the diffractive properties were analyzed on the basis of coupled wave differential equations in which second-order derivatives were neglected. Such an approach is justified for one-dimensional gratings in optical waveguides where the gratings represent weak modulation of the refractive index (its real and/or imaginary part) without any significant changes in its average value in the grating portion of the waveguide. In the case of slab gratings, illustrated in Fig.~1, neglecting the second derivatives of the field amplitudes is equivalent to neglecting the boundary effects, i.e. the bulk diffracted orders are retained while the waves produced at the boundaries are eliminated. Such an approximation could lead to significant errors.  In the case of $PT$-symmetric gratings, where the diffraction modes have a very unusual interaction mechanism, it is very important to study how the slab boundaries affect the diffraction  and  how they affect invisibility in the two-dimensional $PT$-symmetric volume grating.

We have therefore analyzed diffraction from such a slab by using the full, second-order Maxwell equations. In Sec.~\ref{2MS} we study a two-mode solution valid for angles near Bragg incidence. This applies for an arbitrary ratio between the index and gain/loss modulations, allowing us to track properties from standard index grating to a $PT$ symmetric grating at the symmetry-breaking point. Then in Sec.~\ref{analytic} we specialize to this latter grating. Due to the particular directed structure of the coupled equations we are able to derive analytic expressions for the
first three diffractive orders, $S_0$, $S_1$ and $S_2$. In the following sections \ref{FS}-\ref{antisymm} we use these expressions to analyze the properties of the $PT$-grating in a variety of different configurations characterized by the values of the background diffractive index within and on either side of the slab, including a possible reflective layer at the back of the slab. A discussion of the general properties of this type of grating along with our conclusions is given in Sec.~\ref{conc}.

\section{Second-order coupled-mode equations}
 In this paper we study the diffraction characteristics of active holographic gratings as a gain/loss
modulation in combination with traditional index gratings.
\begin{figure}[h!]
\begin{center}
\resizebox{!}{7cm}{\includegraphics{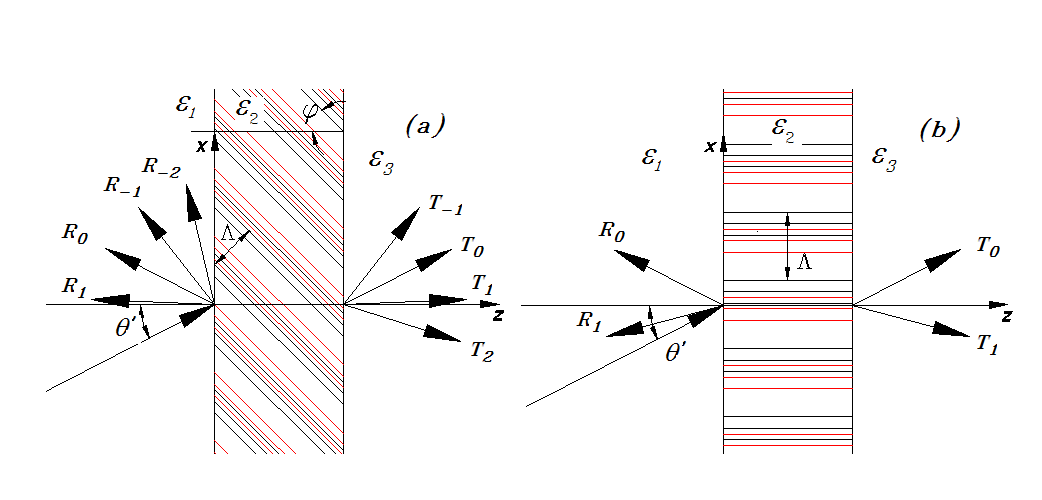}}
\caption{(a) Planar slanted grating of the index (black color fringes) and gain/loss (red color fringes) modulation and
(b) non-slanted grating}
\end{center}
\end{figure}
The slanted grating is assumed to be composed of modulation of the relative dielectric permittivity
\beq
\ve(x,z)=\ve_2+\Delta \ve \cos(K(x\sin\vf +z\cos\vf))
\eeq
and modulation of gain and loss
\beq
\sigma(x,z)=\Delta\sigma \sin(K(x\sin\vf +z\cos\vf))
\eeq
in the region from $z = 0$ to $z = d$ with the same spatial frequency shifted by a quarter of period $\Lambda/4$ ($K=2\pi/\Lambda$)
with respect to one another, where $\ve_2$ is the average relative permittivity in the grating area, $\Delta\ve$ is the amplitude of the sinusoidal relative permittivity, $\Delta\sigma$ is
the amplitude of the gain/loss periodic distribution, and $\vf$ is the grating slant angle. Unlike traditional modulation of the refractive index, Eq.~(2) describes modulation of its imaginary part, so we will call the grating of Eq.~(1) the real grating, and the grating described by Eq.~(2)  the imaginary one.  Fig.~1 shows the generalized model of the hologram grating used in our study. It covers the case of free-space to free-space diffraction as well as planar slab holograms. The propagation constant $k(x,z)$ inside the grating slab is spatially modulated and related to the
relative permittivity $\ve(x,z)$ and the gain/loss distribution $\sigma(x,z)$ by the well-known formula
\beq
k^2(x,z)=k_0^2\ve(x,z)-j\w\mu\sigma(x,z)\ ,
\eeq
where $\mu$ is the permeability of the medium, $\omega$ is the angular frequency of the wave and  $k_0=\omega/c$ is the wave-vector in free space, related to the free-space wavelength $\lambda_0$ by $k_0=2\pi/\lambda_0$.

Equations (1) - (3) can be combined in the following form:
\beq
k^2(x,z)=k_2^2+2k_2\left(\kappa^- \exp(jKr)+\kappa^+ \exp(-jKr)\right)\ ,
\eeq
where $k_2=k_0(\ve_2)^\half$ is the average propagation constant and $r$ is the coordinate vector. The coupling constants
$\kappa^+$ and $\kappa^-$ are
\beq
\kappa^\pm=\frac{1}{4(\ve_2)^\half}\left(k_0\Delta\ve \pm c\mu\Delta\sigma\right)
\eeq
They can take quite different values, unlike the situation with only real or imaginary gratings, where the coupling constants
are always equal, at least in magnitude.

In the two unmodulated regions, $z < 0$ and $z > d$, where we assume uniform permittivity $\ve_1$ and $\ve_3$, respectively,
the assumed solutions of the wave equation for the normalized electric fields are, for $z<0$ (incident and reflected waves):
\bea
E_1(x,z)&=&\exp\left[-jk_1 (x\sin\theta'+z\cos\theta')\right]+\hspace{8cm}\cr
&+& \sum_{m=-\infty}^{\infty} R_m\exp\left[-j\{(k_2\sin\theta-mK\sin\vf)x-(k_1^2-(k_2\sin\theta-mK\sin\vf)^2)^\half z\}\right]
\cr&&
\eea
and for $z>d$ (transmitted waves)
\beq
E_3(x,z)=
\sum_{m=-\infty}^{\infty} T_m\exp\left[-j\{(k_2\sin\theta-mK\sin\vf)x+(k_3^2-(k_2\sin\theta-mK\sin\vf)^2)^\half (z-d)\}\right]
\eeq
The total electric field in the hologram region $0< z < d$   is the superposition of multiple waves:
\beq
E_2(x,z)=
\sum_{m=-\infty}^{\infty} S_m(z)\exp\left[-j(k_2\sin\theta-mK\sin\vf)x\right]\ ,
\eeq
where $k_1=k_0(\ve_1)^\half$, $k_3=k_0(\ve_3)^\half$, $\theta'$ is the angle of incidence in Region 1, and $\theta$ is the angle of refraction in Region 2, related to each other by $k_1\sin\theta'=k_2\sin\theta$. In these equations $R_m$, and $T_m$ are the amplitudes of the $m$-th reflected and transmitted waves and are to be determined. $S_m$(z) is the amplitude of the $m$-th wave in the modulated region and is to be determined by solving the wave equation for an incident plane wave with TE polarization (i.e. electric field perpendicular to
the plane of incidence)
\beq
\nabla^2 E_2(x,z)+k_0^2\ \ve(x,z)E_2(x,z)=0
\eeq
To find $S_m(z)$, Eqs.~(1) and (8) are substituted into Eq.~(9), resulting in the system of coupled-wave equations \cite{JAK,GM1}:
\bea
&&\frac{d^2S_m(z)}{dz^2} +\left[k_2^2-(k_2\sin\theta -mK \sin\vf)^2 \right] S_m(z)+\hspace{6cm}\cr&&\cr
&&\hspace{2cm}+2k_2\left[\kappa^- e^{j K z \cos\vf}S_{m+1}(z)+\kappa^+ e^{-j K z\cos\vf}S_{m-1}(z)\right]=0
\eea
This set of coupled-wave equations contains no first-derivative terms. In addition, Eqs.~(10) are
nonconstant-coefficient differential equations due to the presence of $z$ in the coefficients of the $S_{m-1}$ and $S_{m+1}$ terms.

From now on we will restrict ourselves to the case of an unslanted grating, taking $\vf= \pi/2$.
In this case the fringes are perpendicular to the slab boundaries $z = 0$ and $z =d$, cf. Fig.~1(b), and the equations become constant-coefficient differential equations. 	

For $\theta$ near the (first) Bragg angle $\theta_B$, given by $K=2k_2\sin\theta_B$, only the zeroth-order and the first-order diffraction modes are coupled strongly to each other. Retaining only these two modes, Eqs.~(10) become:
\bea\label{eqs1and2}
\frac{d^2S_0(z)}{dz^2} +k_2^2\cos^2\theta\ S_0(z)+2k_2\kappa^- S_1(z)&=&0\cr&&\\
\frac{d^2S_1(z)}{dz^2} +\left[k_2^2-(k_2\sin\theta-K)^2 \right] S_1(z)+2k_2\kappa^+ S_0(z)&=&0\nonumber
\eea
At the exact Bragg condition, when $\theta=\theta_B$, the coupled equations reduce to the following form in terms of the dimensionless
coordinate $u =k_2z$:
\bea
\frac{d^2S_0(u)}{du^2} +\cos^2\theta_B S_0(u)+\xi_1 S_1(u)&=&0\cr&&\\
\frac{d^2S_1(z)}{dz^2} +\cos^2\theta_B S_1(u)+\xi_2 S_0(u)&=&0\nonumber
\eea
where
\beq
\xi_1= 2\kappa^-/k_2 \hspace{1cm} {\rm and} \hspace{1cm} \xi_2=2\kappa^+/k_2
\eeq
The coupled equations (13) can be decoupled by switching to
\beq
V_0=S_0+\sqrt{\xi_1/\xi_2}\ S_1 \hspace{1cm} {\rm and} \hspace{1cm} V_0=S_0-\sqrt{\xi_1/\xi_2}\ S_1\ ,
\eeq
when the equations become
\bea
\frac{d^2V_0(u)}{du^2}+\rho_1^2V_0(u)&=&0\cr&&\\
\frac{d^2V_1(u)}{du^2}+\rho_2^2V_1(u)&=&0\nonumber
\eea
where
\beq
\rho_1=(\cos^2\theta_B+\sqrt{\xi_1\xi_1})^\half \hspace{1cm} {\rm and} \hspace{1cm} \rho_2=(\cos^2\theta_B-\sqrt{\xi_1\xi_1})^\half \ .
\eeq
Then $S_0$ and $S_1$ are given by $S_0(u)=\half(V_0(u)+V_1(u))$ and $S_1(u)=\half\sqrt{\xi_2/\xi_1}(V_0(u)-V_1(u))$, where $V_0(u)$ and $V_1(u)$ have the solutions
\bea\label{Bragg}
V_0(u)&=&A e^{j\rho_1 u}+B e^{-j\rho_1 u}\cr
&&\\
V_1(u)&=&C e^{j\rho_2 u}+D e^{-j\rho_2 u}\nonumber
\eea
in which the constants $A$, $B$, $C$ and $D$ are to be found from the boundary conditions.

These require that the tangential electric and tangential magnetic fields be continuous across the two
boundaries ($z =0$ and $z = d$). For the $H$-mode polarization discussed in this paper, the electric field only has a component in the
$y$-direction and so it is the tangential electric field directly. The magnetic field intensity, however, must be obtained through the
Maxwell equation. The tangential component of $H$ is in the $x$-direction and is thus given by $H_x=(-j/(\w \mu_0))\p E_y/\p z$.

In the approximation of keeping only the two modes $S_0(u)$ and $S_1(u)$ the
four quantities to be matched and the resulting boundary conditions are\\

\noindent a) tangential $E$ at $z = 0$:
\beq\label{bc1}
1+R_0=S_0(0),\hspace{1cm} R_1=S_1(0)
\eeq
\noindent b) tangential $H$ at $z = 0$ :
\beq\label{bc2}
k_2 S_0'(0)= j (k_1^2-k_2^2 \sin^2\theta)^\half (R_0-1),\hspace{1cm}
k_2 S_1'(0)= j [k_1^2-(k_2\sin\theta -K)^2]^\half R_1
\eeq
\noindent c) tangential $E$ at $z= d$:
\beq\label{bc3}
T_0=S_0(d),\hspace{1cm} T_1=S_1(d)
\eeq
\noindent c) tangential $H$ at $z = d$:
\beq\label{bc4}
k_2 S_0'(d)= -j (k_3^2-k_2^2 \sin^2\theta)^\half T_0,\hspace{1cm}
k_2 S_1'(d)= -j [k_1^2-(k_2\sin\theta -K)^2]^\half T_1
\eeq
\section{Two-mode solution for $\theta=\theta_B$}\label{2MS}
Taking $S_0(u)$ and $S_1(u)$ as given in Eqs.~(\ref{Bragg}), the boundary conditions (\ref{bc1}) - (\ref{bc4})
lead to the following eight equations for the eight unknown constants: $A$, $B$, $C$, $D$, $R_0$, $R_1$, $T_0$ and $T_1$.
\begin{subequations}
\bea
R_0 +\xi R_1+1&=&A+B \\
\alpha_B (R_0+\xi R_1-1)&=&\rho_1(A-B)\\
R_0-\xi R_1+1&=&C+D\\
\alpha_B(R_0-\xi R_1-1)&=&\rho_2(C+D)\\
T_0+\xi T_1&=&Ae^{j\rho_1 u_d}+B e^{-j\rho_1 u_d}\\
T_0-\xi T_1&=&Ce^{j\rho_2 u_d}+D e^{-j\rho_2 u_d}\\
-\beta_B(T_0+\xi T_1)&=&\rho_1(Ae^{j\rho_1 u_d}-B e^{-j\rho_1 u_d})\\
-\beta_B(T_0-\xi T_1)&=&\rho_2(Ce^{j\rho_2 u_d}-D e^{-j\rho_2 u_d})
\eea
\end{subequations}
where $\xi=\surd(\xi_1/\xi_2)$, $\alpha_B=\surd(\ve_1/\ve_2-\sin^2\theta_B)$, $\beta_B=\surd(\ve_3/\ve_2-\sin^2\theta_B)$
and $u_d=k_2 d$.

Solving these equations, we find the following expressions for the zeroth- and first-order reflection
coefficients:
\beq\label{BS1}
R_0=\half(F(\rho_1)+F(\rho_2)), \hspace{2cm} R_1=\frac{1}{2\xi}(F(\rho_1)-F(\rho_2))
\eeq
where, for $m=1,2$,
\beq
F(\rho_m)=\frac{(\rho_m-\beta_B)(\alpha_B+\rho_m)e^{-j\rho_m u_d}+(\rho_m+\beta_B)(\alpha_B-\rho_m)e^{j\rho_m u_d}}{(\rho_m-\beta_B)(\alpha_B-\rho_m)e^{-j\rho_m u_d}+(\rho_m+\beta_B)(\alpha_B+\rho_m)e^{j\rho_m u_d}}
\eeq
The transmission coefficients are expressed in terms of $G(\rho_1)$ and $G(\rho_2)$ as
\beq
T_0=\half(G(\rho_1)+G(\rho_2)), \hspace{2cm} T_1=\frac{1}{2\xi}(G(\rho_1)-G(\rho_2))\ ,
\eeq
where
\beq\label{BS4}
G(\rho_m)=\frac{4\rho_m\alpha_B}{(\rho_m-\beta_B)(\alpha_B-\rho_m)e^{-j\rho_m u_d}+(\rho_m+\beta_B)(\alpha_B+\rho_m)e^{j\rho_m u_d}}
\eeq
It should be clear that variation of the asymmetry coefficient $\xi$ from 1 to 0 describes the transition from a traditional index grating,
analyzed by Kong \cite{JAK}, to a $PT$-symmetric one which reaches its balanced form at $\xi=0$, as shown in Figs.~2(a)-(d) for $\xi=1$  (magenta, dot-dashed curves), $\xi=0.5$ (green, dashed curves), $\xi=0.25$ (blue, dotted curves) and finally the $PT$-symmetric case (red, solid) for the slab with $\ve_2=2.4$ in air, $\ve_1=\ve_3=1$.

Unlike the solution for the $PT$-symmetric grating
obtained through the first-order coupled wave equations \cite{MKBK}, which provides only the transmission coefficients, with
$|T_0|=1$  and $T_1\propto \xi_2 u_d$ when $\theta=\theta_B$, our solution shows significant intensities in the zeroth and first reflective orders. In fact, power redistribution results in a normalized power reduction in $|T_0|^2$ from 1 to 0.83, with $|R_0|^2=0.17$. As expected, the mode-coupling nature in $PT$-symmetric gratings does not provide any amplification for the zeroth orders either in transmission or reflection. However, the first diffraction orders exhibit linear growth in amplitude, quadratic in power, before they reach gain saturation.
\begin{figure}[h!]
\begin{center}
\hspace{-1cm}\resizebox{!}{9cm}{\includegraphics{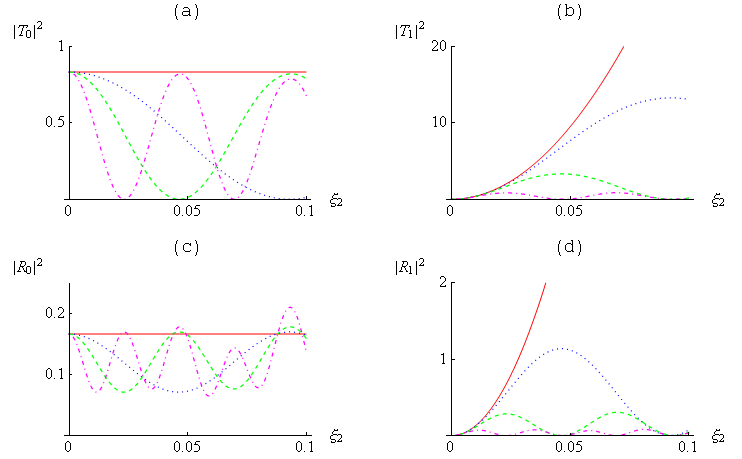}}
\caption{Two-mode solution for incidence at the first Bragg angle $\theta'_B$: transmission and reflection coefficients as functions of the grating strength for different values of $\xi=\sqrt{\xi_1/\xi_2}$, where  $\xi=1$ (magenta, dot-dashed) corresponds to a traditional index grating and $\xi=0$ (red, solid) describes an ideal balanced $PT$-symmetric grating. The other values shown are $\xi=0.5$ (green, dashed) and  $\xi=0.25$ (blue, dotted).The remaining parameters are $\ve_1=1$, $\ve_2=2.4$, $\ve_3=1$, $d$ = 8 $\mu$m, $\Lambda$= 0.5 $\mu$m, $\lambda_0$=0.6328 $\mu$m.}
\end{center}
\end{figure}

The assumptions of neglecting the second derivatives of field amplitudes and neglecting boundary effects transform the problem
into a filled-space problem \cite{GM1}, like a grating filling all space with imaginary boundaries at $z =0$ and $z = d$.
In our second-order derivative solution we can approach such a regime by putting $\ve_1=\ve_2=\ve_3$. Indeed, as we can see in Fig.~3, the zeroth-order transmission amplitude returns to unity (red solid line in Fig.~3(a)) with practically no reflection
(Fig.~3(c)). Reflected light in the first order is also practically negligible (Fig.~3(d)). Note that the power
supplied to $T_1$ comes from the active grating, not at the expense of the zeroth-order diffraction,
which still satisfies $|R_0|^2+|T_0|^2=1$.
\begin{figure}[h!]
\begin{center}
\hspace{-0.8cm}\resizebox{!}{9cm}{\includegraphics{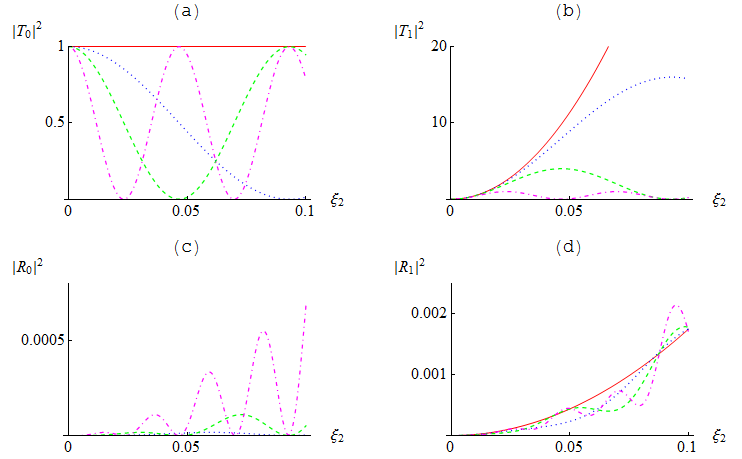}}
\caption{Two-mode solution for incidence at the first Bragg angle $\theta'_B$: transmission and reflection coefficients as functions of the grating strength for the filled-space configuration $\ve_1=\ve_2=\ve_3=2.4$. The remaining quantities are the same as in Fig.~2.}
\end{center}
\end{figure}
\section{Analytic solution for balanced $PT$-symmetric grating for arbitrary angle of incidence}\label{analytic}
The expressions (\ref{BS1})-(\ref{BS4}) for diffraction in transmission
and reflection obtained in the previous section are valid only for $\theta=\theta_B$ but for arbitrary $\xi_1$, $\xi_2$, thus including the perfectly balanced $PT$-symmetric grating as well as the unbalanced one. In fact, these expressions even cover the case of a purely imaginary grating of gain/loss modulation with no index grating in the slab ($\xi_1=-\xi_2$).

In this section we extend our analysis of the balanced $PT$-symmetric grating ($\xi_1$ =0), but with
arbitrary angle of incidence. In that case the coupled wave equations (\ref{eqs1and2}) are
\begin{subequations}
\bea\label{bpa}
&&\frac{d^2S_0(u)}{du^2} +\cos^2\theta\ S_0(u)=0\\ \label{bpb}&&\cr
&&\frac{d^2S_1(u)}{du^2} +\left[1-(2\sin\theta_B-\sin\theta)^2 \right] S_1(u)+\xi_2 S_0(u)=0
\eea
\end{subequations}

The first equation (\ref{bpa}) for the zeroth-order amplitude $S_0(u)$ (non-diffracted light) is decoupled from the second equation for the first-order amplitude $S_1(u)$. We therefore have a solution for  $S_0(u)$  of the form
\beq
S_0(u)= A_0 e^{j u \cos\theta}+B_0 e^{-j u \cos\theta}
\eeq
Applying the boundary conditions (\ref{bc1})-(\ref{bc4}) we can find $T_0$ and $R_0$ and the constants $A_0$ and $B_0$:
\bea\label{T0}
T_0&=& \frac{4 \alpha_0\cos\theta}{(\alpha_0+\cos\theta)(\beta_0+\cos\theta)e^{j u_d \cos\theta} - (\alpha_0-\cos\theta)(\beta_0-\cos\theta)e^{-j u_d \cos\theta}}\\
&&\cr\label{R0}
R_0&=&\frac{(\alpha_0-\cos\theta)(\beta_0+\cos\theta)e^{j u_d \cos\theta} + (\alpha_0+\cos\theta)(\beta_0-\cos\theta)e^{-j u_d \cos\theta}}{(\alpha_0+\cos\theta)(\beta_0+\cos\theta)e^{j u_d \cos\theta} - (\alpha_0-\cos\theta)(\beta_0-\cos\theta)e^{-j u_d \cos\theta}}\\
&&\cr
A_0&=&\frac{T_0}{2}\left(\frac{\cos\theta-\beta_0}{\cos\theta}\right)e^{-j u_d \cos\theta}\hspace{1cm}
B_0=\frac{T_0}{2}\left(\frac{\cos\theta+\beta_0}{\cos\theta}\right)e^{j u_d \cos\theta}
\eea
where $\alpha_0=\surd(\ve_1/\ve_2-\sin^2\theta)$ and $\beta_0=\surd(\ve_3/\ve_2-\sin^2\theta)$.

Eq.~(\ref{bpb}) is an inhomogeneous second-order differential equation for $S_1(u)$, whose solutions can be found as a sum of the general
solution of the homogenous equation, $(S_1)_H$ and a particular solution $(S_1)_I$ of the inhomogeneous equation.
The solution of the homogeneous equation is
\beq
(S_1(u))_H=C_1 e^{j \eta_1 u}+D_1 e^{-j \eta_1 u}
\eeq
where $\eta_1=\surd{[1-(2\sin\theta_B-\sin\theta)^2]}$ and $C_1$ and $D_1$ are constants to be determined. The particular solution can be found using the method of undetermined coefficients. We write
\beq
(S_1(u))_I= A_1 e^{j u \cos\theta} + B_1 e^{-j u \cos\theta}
\eeq
and find that $A_1= x_1 A_0$ and $B_1=x_1 B_0$, where
\beq
x_1=\frac{\xi_2}{4\sin\theta_B(\sin\theta_B -\sin\theta)}
\eeq
Applying the boundary conditions (19)-(22) we can find $T_1$, $R_1$, $C_1$ and $D_1$. The rather lengthy
expressions thus obtained for $T_1$ and $R_1$ can be expressed in a condensed form using the functions\footnote{Note that, using
this notation, the expressions for the zeroth-order transmission and reflection coefficients are simply
$T_0=4 \alpha_0 \eta_0/h(\eta_0,\alpha_0,\beta_0)$ and $R_0 = - h(\eta_0,-\alpha_0,\beta_0)/h(\eta_0,\alpha_0,\beta_0)$.}
\begin{subequations}
\bea
f(a,b,c)&:=&(a+b)(a-c)(1-e^{-j(a-b)u_d})-(a-b)(a+c)(1-e^{j(a+b)u_d})\hspace{1cm}\\
&&\cr
g(a,b,c)&:=&(a+b)(a-c)(e^{-j b u_d}-e^{-j a u_d})-(a-b)(a+c)(e^{-j b u_d}-e^{j a u_d})\hspace{1cm}\\
&&\cr
h(a,b,c)&:=&(a+b)(a+c)e^{j a u_d}-(a-b)(a-c)e^{-j a u_d}\hspace{1cm}
\eea
\end{subequations}
In this notation, and with the definitions $\eta_0 =\cos\theta$, $\alpha_m =\surd{[\ve_1/\ve_2-(2 m \sin\theta_B-\sin\theta)^2]}$ and $\beta_n =\surd{[\ve_3/\ve_2-(2 m \sin\theta_B-\sin\theta)^2]}$,
\bea\label{first order}
T_1&=& \frac{1}{h(\eta_1,\alpha_1, \beta_1)}\left[f(\eta_1, \eta_0,\alpha_1) A_1 + f(\eta_1,-\eta_0,\alpha_1) B_1\right]\\
&&\cr
R_1&=& \frac{1}{h(\eta_1,\alpha_1, \beta_1)}\left[g(\eta_1, -\eta_0,\beta_1) A_1 + g(\eta_1,\eta_0,\beta_1) B_1\right]
\eea
The coefficients $C_1$ and $D_1$ are given by
\bea
C_1&=&\frac{1}{h(\eta_1,\alpha_1, \beta_1)}\left\{ A_1\left[(\alpha_1-\eta_0)(\beta_1-\eta_1)e^{-j\eta_1 u_d}-(\alpha_1+\eta_1)(\beta_1+\eta_0)e^{j\eta_0 u_d}\right]\right. \\
&& \hspace{2.2cm}+\left. B_1\left[(\alpha_1+\eta_0)(\beta_1-\eta_1)e^{-j\eta_1 u_d}-(\alpha_1+\eta_1)(\beta_1-\eta_0)e^{-j\eta_0 u_d}\right]\right\}\nonumber
\eea
and
\bea
D_1&=&\frac{-1}{h(\eta_1,\alpha_1, \beta_1)}\left\{ A_1\left[(\alpha_1-\eta_0)(\beta_1+\eta_1)e^{j\eta_1 u_d}-(\alpha_1-\eta_1)(\beta_1+\eta_0)e^{j\eta_0 u_d}\right]\right. \\
&&\hspace{2.2cm}+\left. B_1\left[(\alpha_1+\eta_0)(\beta_1+\eta_1)e^{-j\eta_1 u_d}-(\alpha_1-\eta_1)(\beta_1-\eta_0)e^{-j\eta_0 u_d}\right]\right\}\nonumber
\eea

It is important to emphasize that the mode coupling in a $PT$-symmetric grating has a unidirectional nature, with energy flowing from lower order to higher order modes: from zeroth order to first order, from first order to second order and so on. With such a type of
coupling it is relatively easy to find practically any higher diffraction order analytically. Here we exploit this feature to derive explicit expressions for the second-order reflection and transmission coefficients.

The equation for the second-order mode has the following form:
\beq
\frac{d^2S_2(u)}{du^2} +\left[1-(4\sin\theta_B-\sin\theta)^2 \right] S_2(u)+\xi_2 S_1(u)=0
\eeq
Using the same approach as to Eq.~(\ref{bpb}), the solution is again sought as a sum of the general solution of the homogeneous
equation and a particular solution of the nonhomogeneous equation.
The solution of the homogeneous equation is
\beq
(S_2(u))_H=E_2 e^{j u\eta_2 }+F_2 e^{-j u\eta_2 }
\eeq
where $\eta_2=\surd{[1-(4\sin\theta_B-\sin\theta)^2]}$ and $E_2$ and $F_2$ are constants to be determined.
The particular solution of the differential equation can again be found using method of
undetermined coefficients. Writing
\beq
(S_2(u))_I=C_2 e^{j u \eta_1}+D_2 e^{-j u \eta_1}+A_2 e^{j u \cos\theta}+A_2 e^{-j u \cos\theta}\ ,
\eeq
we find $C_2=x_3 C_1$,\ $D_3=x_2 D_1$,\ $A_2 = x_2 A_1$ and $B_2=x_2 B_1$,
where
\beq
x_2=\frac{\xi_2}{8\sin\theta_B(2\sin\theta_B -\sin\theta)},
\hspace{1cm} x_3=\frac{\xi_2}{4\sin\theta_B(3\sin\theta_B -\sin\theta)}
\eeq
Applying the boundary conditions at $u=0$ and $u=u_d$ we can find $T_2$, $R_2$, $E_2$ and $F_2$. With the help of the functions
defined in Section \ref{analytic}, we can express $T_2$ and $R_2$ in a rather compact form:
\bea
T_2&=& \frac{1}{h(\eta_2,\alpha_2,\beta_2)}\left[ f(\eta_2,\eta_0,\alpha_2)A_2 + f(\eta_2,-\eta_0,\alpha_2)B_2 +f(\eta_2,\eta_1,\alpha_2)C_2 + f(\eta_2,-\eta_1,\alpha_2)D_2 \right]\cr
&&\cr&&\\
R_2&=& \frac{1}{h(\eta_2,\alpha_2,\beta_2)}\left[ g(\eta_2,\eta_0,\alpha_2)A_2 + g(\eta_2,-\eta_0,\alpha_2)B_2 +g(\eta_2,\eta_1,\alpha_2)C_2 + g(\eta_2,-\eta_1,\alpha_2)D_2 \right]\cr
&&\cr&&
\eea
Similar expressions can be obtained for $E_2$ and $F_2$, but are not needed for our present purposes. They would be needed for the calculation of the third-order reflection and transmission coefficients.

In subsequent figures we will display the diffraction efficiencies rather than the squared moduli of the diffraction coefficients. The diffraction efficiency for the $i$th order is defined as the diffracted intensity of this order divided by the input intensity. We normalized the amplitude of the incident plane wave to one. The diffraction intensities in Regions 1 and 3 are therefore
\bea
&&DER_m={\rm Re}\left(\frac{\alpha_m}{\alpha_0}\right)|R_m|^2
\hspace{1.8cm}
DET_m={\rm Re}\left(\frac{\beta_m}{\alpha_0}\right)|T_m|^2
\eea

\section{Filled-space $PT$-symmetric grating}\label{FS}
As first check of our solution we will consider the particular case of the so-called filled-space grating, when the dielectric permittivity to the left and right of the slab is equal to the average dielectric permittivity of the slab: $\ve_1=\ve_2=\ve_3$. This configuration should provide a solution that is very close to that of the first-order coupled wave equations.

Indeed, when $\ve_1=\ve_2=\ve_3$, then $\alpha_0=\beta_0=\cos\theta$, so that $A_0=0$ and $B_0=1$, $R_0=0$ and $T_0=e^{-j u_d \cos\theta}$. With no reflections from the slab boundaries
the non-diffracted wave passes through the slab without any attenuation/amplification and without any phase modulation, in accordance
with the invisibility property.

On the other hand, the first-order diffraction occurs with strong amplification,
as is seen from Fig.~4(a). For $\ve_1=\ve_2=\ve_3$ the expressions for $T_1$ and $R_1$ in Eqs.~(\ref{first order}) simplify to
\bea
T_1&=&x_1\frac{(\eta_1+\eta_0)}{2\eta_1}(e^{-j \eta_1 u_d}-e^{-j \eta_0 u_d})= \xi_2 \frac{(\eta_1+\cos\theta)(e^{-j \eta_1 u_d}-e^{-j \eta_0 u_d})}{8 \eta_1 \sin\theta_B(\sin\theta_B-\sin\theta)}\\ &&\cr
R_1&=& x_1 \frac{\eta_1-\eta_0}{2\eta_1}(1-e^{-j (\eta_0 +\eta_1)u_d})= \xi_2 \frac{(\eta_1-\cos\theta)(1-e^{-j (\eta_0 +\eta_1)u_d})}{8 \eta_1 \sin\theta_B(\sin\theta_B-\sin\theta)}
\eea
These peak at the Bragg angle, where their values are
\bea
T_1&=&-j\ \frac{\xi_2 u_d}{2\cos\theta_B}e^{-j u_d \cos\theta_B} \cr
&&\\
R_1&=& -j\, \xi_2 \frac{\sin(u_d\cos\theta_B) }{2 \cos^2\theta_B} e^{-j u_d \cos\theta_B}\nonumber                                                                    \eea
The first-order diffraction amplitude $T_1$ grows linearly with the grating strength $\xi_2 u_d$, with amplification close to 800\%
for the parameters chosen in Fig.~4.  This linear growth in amplitude is a characteristic of $PT$-symmetric structures at their breaking point. We should remember that the $PT$-symmetric grating is an active structure: even though the average gain/loss is zero, external energy must be supplied to provide its functionality. $R_1$  is not zero, but is small  even at the resonance, with a diffraction efficiency of less than 0.1\%.
\begin{figure}[h!]
\begin{center}
\hspace{-0.5cm}\resizebox{!}{5.3cm}{\includegraphics{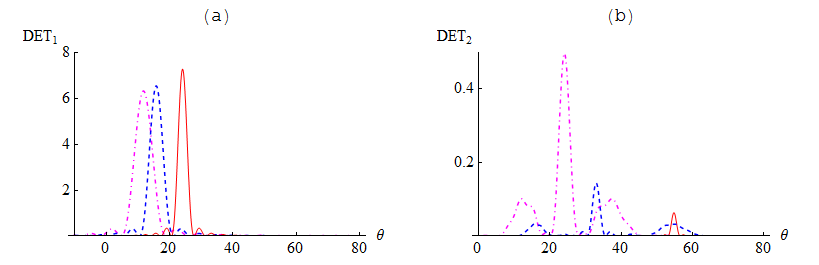}}
\caption{Filled-space configuration ($\ve_1=\ve_2=\ve_3=2.4$) : diffraction efficiency in (a) first and (b) second orders in transmission as a function of the internal angle of incidence for $\Lambda$= 0.5 $\mu$m (red, solid), $\Lambda$= 0.75 $\mu$m (blue, dashed), $\Lambda$= 1.0 $\mu$m (magenta, dot-dashed). The other parameters are $d$ = 8 $\mu$m,  $\lambda_0$=0.633 $\mu$m.}
\end{center}
\end{figure}

\section{Symmetric Slab Configuration}\label{symm}
In this section we compare the transmission and reflection characteristics of the filled-space
$PT$-symmetric grating without reflections from the slab boundaries ($\ve_1=\ve_2=\ve_3=2.4$)  and the real configuration of the slab in air ($\ve_1=\ve_3=1$, $\ve_2=2.4$). The results are presented in Fig.~5.

\begin{figure}[h!]
\begin{center}
\hspace{-1cm}\resizebox{!}{6.8cm}{\includegraphics{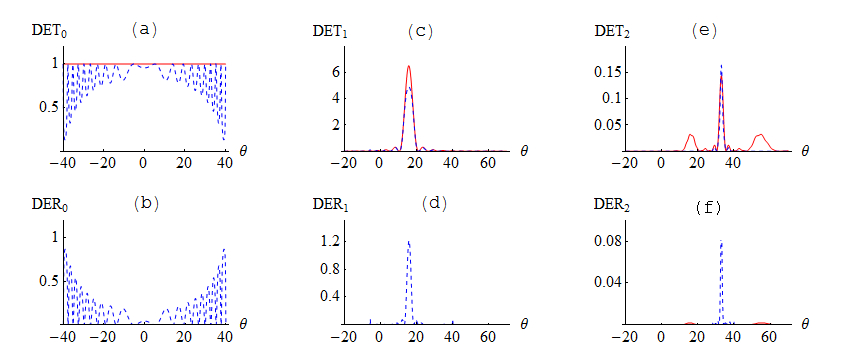}}
\caption{Symmetric vs. filled-space configuration: diffraction efficiency in transmission ((a), (c), (e)) and reflection ((b), (d), (f)) as a function of the internal angle of incidence $\theta$  for  $\ve_1=\ve_3=1$, $\ve_2=2.4$ (blue, dashed), $\ve_1=\ve_2=\ve_3=2.4$ (red, solid). The other parameters are $d$ = 8$\mu$m,  $\Lambda$= 0.75 $\mu$m and $\lambda_0$=0.633 $\mu$m.}
\end{center}
\end{figure}
The reflections from the front and back surfaces of the slab significantly change the spectral characteristics of the zeroth-order  transmission (Fig.~5(a)) and reflection (Fig.~5(b)). These two plots cover the range $-41^{\rm o}<\theta<41^{\rm o}$ for the internal incident angle ($\arcsin(\sqrt{\ve_1/\ve_2})=40.2^{\rm o}$), which corresponds to the range $-90^{\rm o}< \theta' <90^{\rm o}$ for the external incident
angle. The effect of invisibility of the $PT$-symmetric grating for zeroth-order transmission (red solid horizontal line in Fig.~5(a)) is strongly distorted by strong interference of the light reflected from the slab surfaces, as shown by the blue dashed curve. The effect is stronger for larger incident angles. Similarly the zeroth-order reflected light emerges with increasing intensity for larger incident angles (Fig.~5(b)).

The angular spectra for the first-order diffracted light are presented in Fig.~5(c) in transmission and Fig.~5(d) in reflection. As can be seen, the reflection from the slab boundaries leads to a significant increase in the reflected first diffraction order (Fig.~5(d)) along with a rather small decrease of the transmitted light in that order (Fig.~5(c).

\section{Asymmetric Configurations}\label{antisymm}
 In many practical applications the slab supporting the $PT$-symmetric grating might be very thin and fragile and need to be
 attached to a substrate. Such a situation leads to different dielectric permittivity from the left and right sides of the slab, $\ve_1 \ne \ve_3$ . Such a practical requirement might result in the input light incident from the substrate side or from the air side, as shown in
Fig.~6(a) and Fig.~6(b) respectively.
\begin{figure}[h!]
\begin{center}
\resizebox{!}{7cm}{\includegraphics{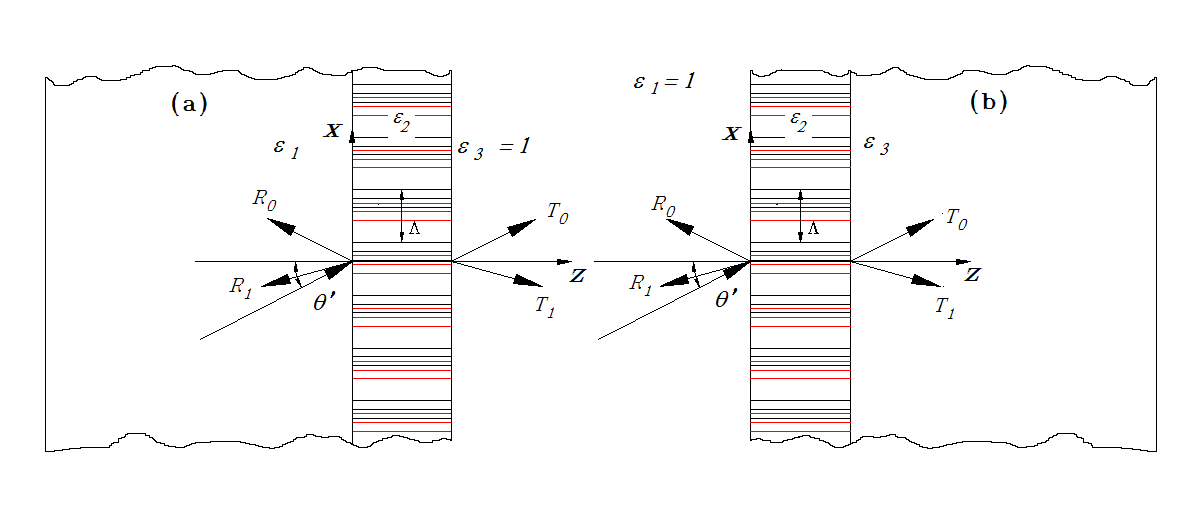}}
\caption{Asymmetric configurations when the input light comes from (a) the substrate side and (b) the air side.}
\end{center}
\end{figure}
\subsection{Light incident from the substrate side: $\ve_3=1$} The geometry presented in Fig.~6(a) has been analyzed, with the results shown in Fig.~7.
\begin{figure}[h!]
\begin{center}
\resizebox{!}{12cm}{\includegraphics{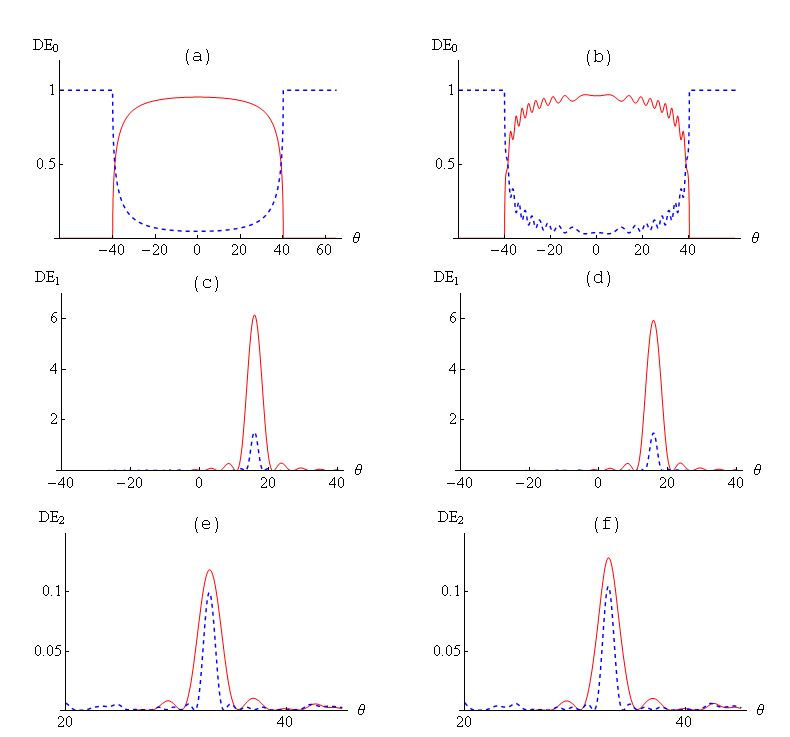}}
\caption{Light incident from substrate: transmission (red, solid) and reflection (blue, dashed) angular spectra for zeroth order (a) and (b), first-order (c) and (d) and second-order light (e) and (f) as functions of the internal angle of incidence.  In the left-hand panels $\ve_1=\ve_2=2.4$, while in the right-hand panels $\ve_1=2.0$, $\ve_2=2.4$. The remaining parameters are  $\ve_3=1$, $d$ = 8 $\mu$m;  $\Lambda$ = 0.75 $\mu$m; $\lambda$=0.633 $\mu$m and $\xi=0.04$. }
\end{center}
\end{figure}

We consider the cases when the permittivity of the substrate and the average permittivity of the slab are the same, $\ve_1=\ve_2$, (Figs.~7(a), (c) and (e)), and when they are different (Figs.~7(b), (d) and (f)). Comparing Figs.~7(a), (b) with Figs.~5(a), (c)
one can see a significant difference in the angular spectral behavior in zeroth order.
Equations (\ref{T0}) and (\ref{R0}) simplify significantly for $\ve_1=\ve_2$, when $\alpha_0=\cos\theta$, so that
\bea
T_0&=&\left(\frac{2\cos\theta}{\cos\theta+\beta_0}\right)e^{-j u_d \cos\theta}\\
&&\cr
R_0&=&\left(\frac{\cos\theta-\beta_0}{\cos\theta+\beta_0}\right)e^{-2j u_d \cos\theta},
\eea
which are  basically the Fresnel coefficients. The transmission rapidly decreases to zero for $|\theta|>\theta_{TIR}\equiv \arcsin(\surd(\ve_3/\ve_2))$, the angle at which total internal reflection occurs at the second surface.  Reflection in zeroth order is close to zero (4.6\%), ($R_0=(\sqrt{\ve_2}-\sqrt{\ve_3})/(\sqrt{\ve_2}+\sqrt{\ve_3})$), for normal incidence and then rapidly increases to 100\% for $|\theta|>\theta_{TIR}$. Introduction of the second reflective
interface, (Fig.7(b)), produces a weak rippling effect on the transmission and reflection spectra.

There is no significant difference in transmission and reflection of the first and second diffraction orders  between the configuration of Figs.~7(a), (c) and (e)  and that of Figs.~7(b), (d) and (f). It seems clear that it is reflection from the interface between the slab and Region 3 that produces the major contribution to the reflective diffraction. If that is the case, then intuitively the reflective diffraction orders can be significantly reduced by illuminating from the air, as shown in Fig.~6(b).

\subsection{ Light incident from the air: $\ve_1=1$}
Indeed, in this set-up the reflection is practically invisible (blue dashed curves) in Fig.~8 for the first and second
diffractive orders. Even the reflection from the slab-substrate interface where $\ve_2 \ne \ve_3$ does not contribute in any significant way to the reflective diffraction orders, Fig.~8(b) and Fig.~8(d). The second-order diffraction is also negligible compared to the first order, so that practically all the light diffracted by the $PT$-symmetric volume grating goes into the first transmissive  diffraction order.
\begin{figure}[h!]
\begin{center}
\resizebox{!}{9cm}{\includegraphics{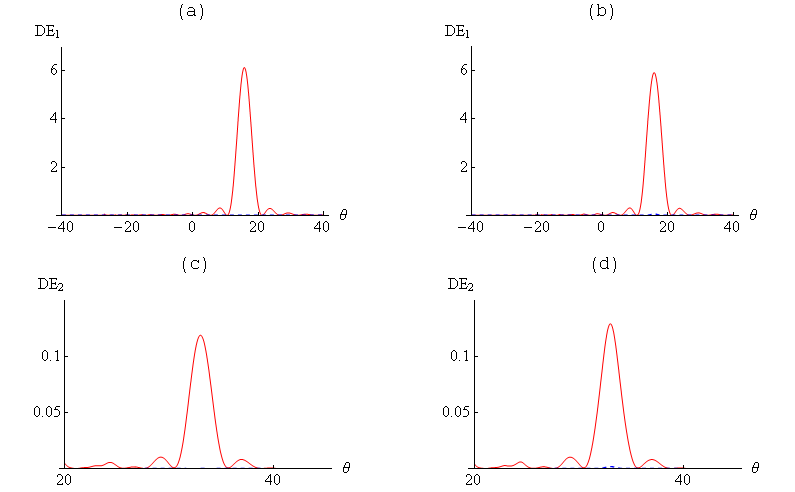}}
\caption{Light incident from the air: transmission (red, solid) and reflection (blue, dashed) angular spectra for first-order (a) and (b) and second-order diffracted light (c) and (d) as functions of the internal angle of incidence. In the left-hand panels $\ve_2=\ve_3=2.4$, while in the right-hand panels $\ve_2=2.4$, $\ve_3=2.0$. The remaining parameters are  $\ve_1=1$, $d$ = 8 $\mu$m;  $\Lambda$ = 0.75 $\mu$m; $\lambda$=0.633 $\mu$m and $\xi=0.04$. }
\end{center}
\end{figure}

\subsection{ Reflective set-up}
To conclude this analysis of the $PT$-symmetric transmission grating we propose a method to reverse its first transmission order into reflection. This can be done by placing an aluminum layer between the slab and the substrate. If this aluminum layer is of the order of one micron in thickness then any influence of the substrate will be shielded. Such a structure can be accurately simulated by assigning the dielectric permittivity of Region 3 the aluminum permittivity at  $\lambda_0$=0.633 $\mu$m, namely $\ve_3=-54.705+21.829 j$. The results are depicted in Fig.~9.  As  can be seen, the reflection now becomes dominant in zeroth order, as well as for first-order diffraction. The peak in the reflection spectrum (dashed blue curve) is now at least an order of magnitude stronger than that in the transmission spectrum (solid red curve.

\begin{figure}[h!]
\begin{center}
\resizebox{!}{5cm}{\includegraphics{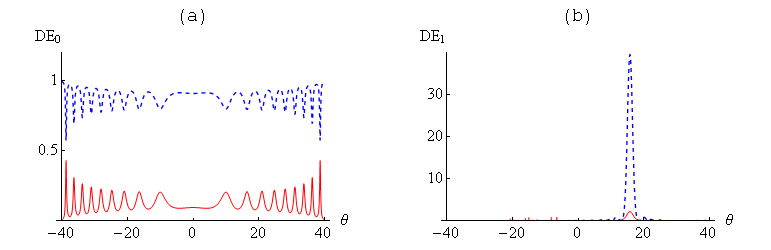}}
\caption{Reflective set-up ($\ve_1=1.0$, $\ve_2=2.4$, $\ve_3=-54.7+21.83 j$): transmission (red, solid) and reflection (blue, dashed)  angular spectra for zeroth-order (a) and first-order diffracted light (c) as functions of the internal angle of incidence for  $d$ = 8 $\mu$m,  $\Lambda$ = 0.75 $\mu$m, $\lambda_0$=0.633 $\mu$m, $\xi=0.004$.}
\end{center}
\end{figure}

\section{Discussion}\label{conc}

In a normal (index) grating, the situation is symmetric between $\theta$ and $-\theta$. Thus, near $\theta=\theta_B$, the first two modes excited are $S_0$ and $S_1$, while near $\theta=-\theta_B$, the first two modes excited are $S_0$ and $S_{-1}$. More generally $\theta \leftrightarrow -\theta$ corresponds to $m \leftrightarrow -m$, where $m$ labels the diffraction order.

However, in a balanced $PT$ grating, this symmetry is lost because the index modulation has an inbuilt direction (it is symmetric under $PT$, but not under $P$ itself). So in the situation we have been describing in the bulk of the paper, illustrated in Fig.~10(a), light incident near the first Bragg angle produces strong signals in first-order diffraction, particularly in transmission. In contrast, for incidence at $\theta$ near $-\theta_B$, as illustrated in Fig.~10(b), there is  essentially no diffraction.

The $PT$ grating is also left-right asymmetric, even when $\ve_1=\ve_3$. In Fig.~10(c), light incident in the reverse direction
of the transmitted beam in Fig.~(a) does not produce the mirror-image of Fig.~10(a), but rather that of Fig.~10(b). Likewise for Fig.~10(d), which is the mirror-image of Fig.~10(a) rather than Fig.~10(b).

There is yet another type of asymmetry of the $PT$ grating. We have called the grating ``balanced" when the perturbation of the refractive index is $\Delta\tilde{n}=\Delta n_0 e^{2\pi j z/\Lambda}$, resulting in $\xi_1=0$ and the consequences explored in the paper. However, if the phase of the gain/loss modulation relative  to the index modulation is reversed, $\Delta\tilde{n}$ instead becomes $\Delta\tilde{n}=\Delta n_0 e^{-2\pi j z/\Lambda}$ and the roles of $\xi_1$ and $\xi_2$ are interchanged, so that now $\xi_2=0$.
In that case the first mode to be excited is $m=-1$, and the coupled equations for $S_0$ and $S_{-1}$ become
\begin{subequations}
\bea\label{bpachangesign}
&&\frac{d^2S_0(u)}{du^2} +\cos^2\theta\ S_0(u)=0\\&&\cr
&&\frac{d^2S_{-1}(u)}{du^2} +\left[1-(2\sin\theta_B+\sin\theta)^2 \right] S_{-1}(u)+\xi_1 S_0(u)=0,
\eea
giving a strong excitation of $S_{-1}$ near $\theta=-\theta_B$. Thus the symmetry $\theta\to -\theta$ of a normal index grating is regained
\emph{provided} that the phase of the gain/loss modulation is reversed at the same time.
\end{subequations}
\begin{figure}[h!]
\begin{center}
\resizebox{!}{7.3cm}{\includegraphics{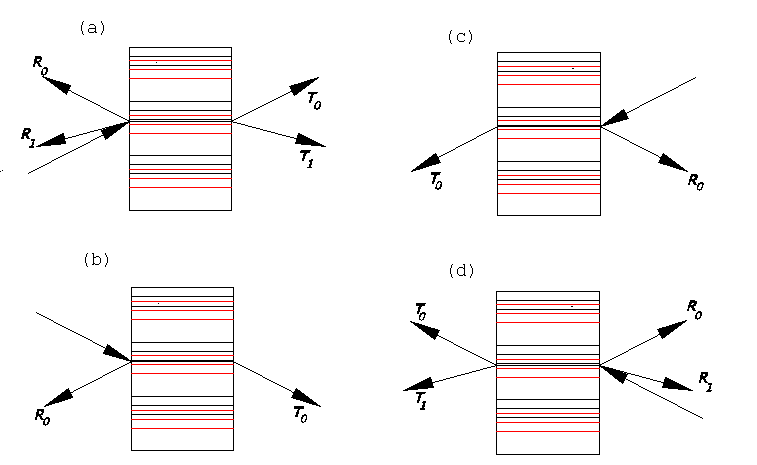}}
\caption{Prominent modes of the $PT$-symmetric grating for incidence at different angles and from different sides:
(a) from the left near the first Bragg angle $\theta_B$; (b) from the left near -$\theta_B$; (c) from the right near $\theta_B$; (d) from the right near -$\theta_B$.  }
\end{center}
\end{figure}

Another prominent characteristic of a balanced $PT$ grating in the paraxial approximation is invisibility \cite{MVB, MKBK}, that is to say that the
transmission coefficient $T_0$ of the undiffracted wave is unity. However, this approximation cannot account for reflections at the boundaries. The main part of our paper has been to find to analytic solutions of the full second-derivative equations for the first three diffractive orders. With the help of these solutions we have analyzed diffraction from the slab in a variety of different configurations.
The invisibility property has been shown to hold only in the filled-space situation, when the background refractive indices are the same, but when this is not the case the reflections produced by the second-order equations result in a significant reduction of $|T_0|^2$.
The linear rise with grating strength of the first-order transmission amplitude, already seen in paraxial approximation, persists when the full second-order equations are used, making $T_1$ by far the strongest signal for the range of parameters we considered.

In Sections \ref{FS}, \ref{symm} and \ref{antisymm} we considered a variety of configurations of the slab in terms of the different background refractive indices $\ve_1$, $\ve_2$, $\ve_3$, showing in detail how the transmitted and reflected light was affected by these different parameters. In the last subsection of Sec.~\ref{antisymm} we showed how a reflective layer at the back of the slab could turn it into a reflective grating, with a strong reflection coefficient $R_1$.

A $PT$-symmetric volume grating is a structure with many interesting and unusual properties, which can only be fully analyzed using the second-order Maxwell equations that we have treated here.



\begin{thebibliography}{99}
\bibitem {Pal}
L.~Paladian, ``Resonance mode expansions and exact solutions for nonuniform gratings," Phys.~Rev.~E, {\bf 54}, 2963-2975
(1996).
\bibitem{MK} M.~Kulishov, J.~M.~Laniel, N.~B\'elanger, J.~Aza\~na, D.~V.~Plant, ``Nonreciprocal waveguide Bragg gratings", Optics Express,
{\bf 13}, 3068-3078 (2005).
\bibitem{GO} M.~Greenberg, M.~Orenstein, ``Irreversible coupling by use of dissipative optics," Opt.~Lett., {\bf 29}, 451-453 (2004).
\bibitem{SL} S.~Longhi, ``Invisibility in $PT$-symmetric complex crystals", J.~Phys.~A {\bf 44}, 485302 (2011)
\bibitem{HFJ}H.~F.~Jones, ``Analytic results for a $PT$-symmetric optical structure", J.~Phys.~A, {\bf 45}, 135306 (2012)
\bibitem{MVB} M.~V.~Berry, ``Lop-sided diffraction by absorbing
crystals," J.~Phys.~Math.~Gen., {\bf 31}, 3493-3502 (1998).
\bibitem{MKBK} M.~Kulishov, B.~Kress, ``Free space diffraction on active gratings with balanced
phase and gain/loss modulations",  Optics Express, {\bf 20}, 29319-29328 (2012).
\bibitem{Feng} L.~Feng, X.~Zhu, S.~Yang, H.~Zhu, P.~Zhang, X.~Yin, Y.~Wang, X.~Zhang, ``Demonstration  of a large-scale optical exceptional point structure," Optics Express, {\bf 22}, 1760-1767 (2014).
\bibitem{BKMK} B.~Kress, M.~Kulishov, ``Parity-time symmetry diffractives implementing unidirectional diffraction - applications
 to optical combiners", Proc. SPIE 9202, 92020F (2014).
\bibitem{JAK} J.~A.~Kong, ``Second-order coupled-mode equations for spatially periodic media", J.~Opt.~Soc.~Am., {\bf 67}, 825-829 (1976).
\bibitem{GM1} T.~K.~Gaylord, M.~G.~Moharam, ``Planar dielectric grating diffraction theories", Appl.~Phys.~B, {\bf 28}, 1-14 (1982).
\end{thebibliography}
\end{document}